\documentclass[conference,compsoc, 10pt]{IEEEtran}
\usepackage{setspace}
\usepackage{color,soul}

\ifCLASSOPTIONcompsoc
  \usepackage{cite}

\usepackage{caption}
\usepackage[pdftex]{graphicx}
\DeclareGraphicsExtensions{.pdf,.jpeg,.png}

\usepackage{booktabs, multirow}
\usepackage[para,online,flushleft]{threeparttable}
\newcommand{\ra}[1]{\renewcommand{\arraystretch}{#1}}

\usepackage{amsmath}
\usepackage{stfloats}
\usepackage[inline]{enumitem}

\usepackage{url}

\begin{document}
%
\title{Exploring Code Clones \\in Programmable Logic Controller Software}


\author{\IEEEauthorblockN{%
  Hannes Thaller\IEEEauthorrefmark{1},
  Rudolf Ramler\IEEEauthorrefmark{2},
  Josef Pichler\IEEEauthorrefmark{2} and
  Alexander Egyed\IEEEauthorrefmark{1}
  }
  \IEEEauthorblockA{
  \IEEEauthorrefmark{1}Institute for Software Systems Engineering, Johannes Kepler University Linz, Austria\\
  Email: hannes.thaller@jku.at, alexander.egyed@jku.at
  }
  \IEEEauthorblockA{
  \IEEEauthorrefmark{2}Software Competence Center Hagenberg GmbH, Austria\\
  Email: rudolf.ramler@scch.at, josef.pichler@scch.at
  }
}

\IEEEoverridecommandlockouts

\maketitle

\begin{abstract}
The reuse of code fragments by copying and pasting is widely practiced in software development and results in code clones.
Cloning is considered an anti-pattern as it negatively affects program correctness and increases maintenance efforts.
Programmable Logic Controller (PLC) software is no exception in the code clone discussion as reuse in development and maintenance is frequently achieved through copy, paste, and modification.
Even though the presence of code clones may not necessary be a problem per se, it is important to detect, track and manage clones as the software system evolves.
Unfortunately, tool support for clone detection and management is not commonly available for PLC software systems or limited to generic tools with a reduced set of features.
In this paper, we investigate code clones in a real-world PLC software system based on IEC 61131-3 Structured Text and C/C++.
We extended a widely used tool for clone detection with normalization support.
Furthermore, we evaluated the different types and natures of code clones in the studied system and their relevance for refactoring.
Results shed light on the applicability and usefulness of clone detection in the context of industrial automation systems and it demonstrates the benefit of adapting detection and management tools for IEC 611313-3 languages.
\end{abstract}


%
\IEEEpeerreviewmaketitle

\section{Introduction}

The increasing share of software for Programmable Logic Controllers (PLCs) and its practical importance have recently been acknowledged by the authors of the 2016 ranking of programming languages in IEEE Spectrum \cite{Cass2016}.
They found that languages for PLCs are on the rise, although in contrast to general-purpose languages such as C, C++ or Java, they specialize in a niche.
Yet their ``relative popularity indicates just how big that niche really is'' \cite{Cass2016}.
With the growing importance of PLC software, an increasing demand for software engineering best practices and tool support is essential.
The focus of this paper is on detecting and analyzing code clones in PLC programs.

Code clones are source code fragments that have been duplicated for reuse, e.g., by copying and pasting \cite{KumarRoy2007}.
Code clones have the reputation to negatively affect program correctness \cite{Juergens2009} and to increase maintenance efforts \cite{KumarRoy2007}.
This form of reuse is widely considered an anti-pattern in software development \cite{Brown1998} and clones are treated as a bad smell in code \cite{Fowler1999}.
Recent studies have shown that there are various reasons why code clones are introduced and that the presence of clones is not per se a problem.
However, the ability to detect, track and manage clones as the software system evolves is of the essence in successful software development \cite{Koschke2008}.

PLC systems are no exception in the discussion of clones as reuse in the industrial automation domain is often achieved through cloning and modifying of existing software systems or sub-systems\cite{Vogel-Heuser2015a}.
This is caused by technological restrictions introduced by programming languages such as the lack of inheritance or polymorphism, organizational limits like time constraints, or simply by the system's complexity.
Furthermore, cloning is often used as a lightweight software product line strategy to cope with various hardware options and application environments.

A wide range of tools and techniques for clone detection and management is available for programming languages such as C, C++ or Java \cite{Bellon2007}.
Similar support for \mbox{IEC 61131-3} languages is mostly limited to general purpose tools, which lack analysis features that require the interpretation of the syntactical structure of the analyzed language.
The contributions of this paper are as follows:
\begin{itemize}
\item Quantitative results from a code clone analysis in a real-world PLC software system consisting of \mbox{IEC 61131-3} Structured Text (ST) and C/C++.
\item The extension of the widely used clone detector Simian \cite{Harris2003a} with language support for Structured Text.
\item A comprehensive study on the relevance and natures of the found clones.
\item An assessment of whether language support in clone detectors is of importance or not.
\end{itemize}

To the best of our knowledge, this paper presents the first study on clone detection for Structured Text.
It sheds light on the questions about the applicability and usefulness of clone detection in the context of industrial automation systems and discusses the need for adapting tools for analyzing programs based on IEC 61131-3 languages.

Section 2 describes the background and related work on clone detection.
The industry context is presented in Section \ref{sec:industry context}, the results of the code clone analysis in Section \ref{sec:code clone analysis}.
Section \ref{sec:code clone study} presents the study on the relevance of found clones, their nature and empirical results on whether detectors should be adapted to support {IEC 61131-3} languages.

\section{Background and Related Work}\label{sec:background}

\begin{figure*}[!t]
	\centering
	\includegraphics[width=\textwidth]{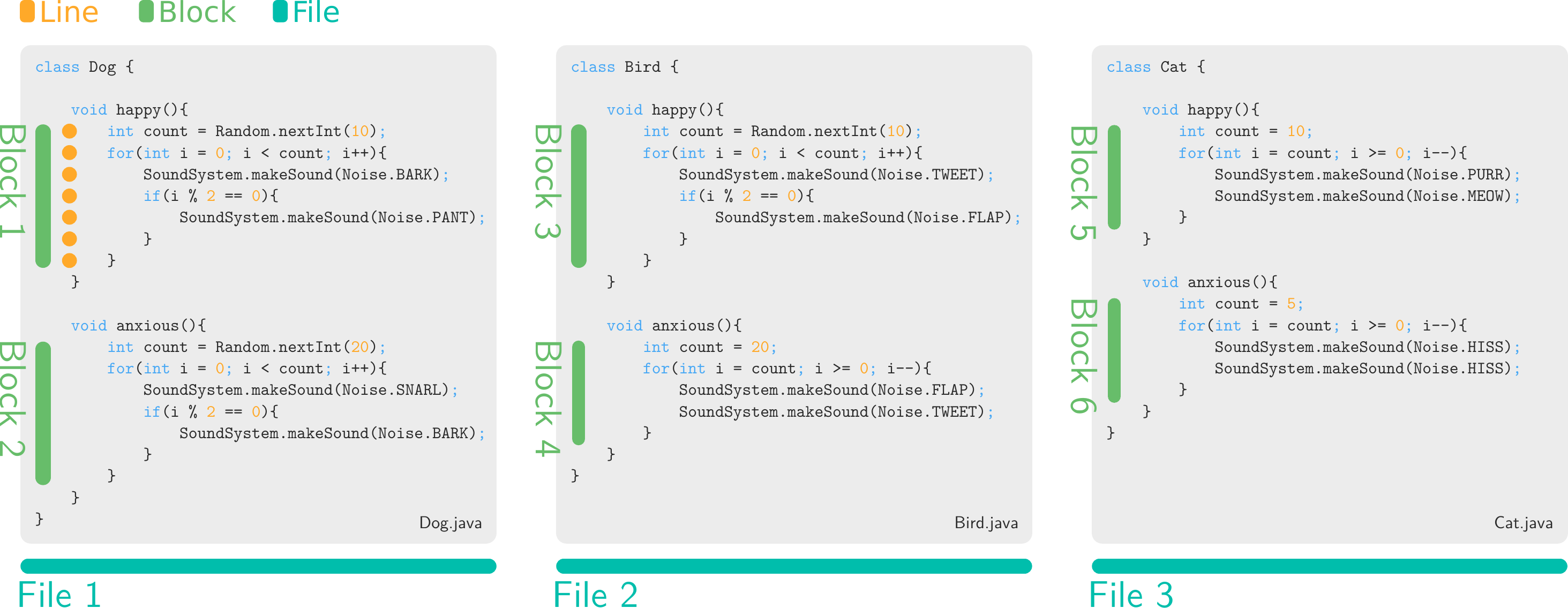}
	\caption{
    Clones may be within a file but also between files and all similar blocks form a clone class (e.g., Block 1-3).
	}
	\label{fig:clone background}
\end{figure*}
Duplicating code fragments during software development activities has a long history.
Definitions, taxonomies, tools for detecting, analyzing, visualizing and managing code clones exist for several languages and technologies.
The interest in code clones is also reflected in the wealth of existing research and the widespread use of tools and techniques in quality management and continuous integration.

\textit{Clone pairs} and \textit{clone classes} \cite{KumarRoy2007} are basic terms used in the context of clone detection.
A clone pair describes two code blocks, also called fragments, that are equal according to a similarity operator.
A clone class is the set of all blocks that are equal according to a similarity operator, effectively forming an equivalence class.
Figure \ref{fig:clone background} shows an example for a clone pair formed by Block 1 and Block 2, where the similarity operator ignores literals and constants.
Examples for clones classes are Block 1-3 and Block 4-6, as they all contain equivalent blocks.

Despite these basic notions, no single holistic definition exists for code clones.
This is due to the different tools and their associated publications that redefine code clones according to the capabilities of the respective tool.
The issue is that code clones are intuitively well-understood, but hard to formalize such that a clear and consistent definition, that covers all applications, has not been found yet.
This study relies on the definition by Baxter et al. \cite{Baxter} as it abstracts detection method specifics without being too vague:
\textit{Code clones are segments of code that are similar according to some definition of similarity.}

\subsection{Clone Taxonomy}\label{sec:clone taxonomy}
A taxonomy based approach helps to align the understanding of code clones, complementary to existing definitions.
The typical and most frequently used categorization of code clones \cite{Bellon2007, Rattan2013, KumarRoy2007} is:
\begin{description}
	\item[Type 1 (Exact Clones):] Program fragments that are identical except for variations in whitespace and comments.
	\item[Type 2 (Parameterized Clones):] Program fragments that are structurally/syntactically similar except for changes in identifiers, literals, types, layout and comments.
	\item[Type 3 (Near-Miss Clones):] Program fragments that include insertions or deletions in addition to changes in identifiers, literals, types and layouts.
	\item[Type 4 (Semantic Clones):] Program fragments that are functionally similar (i.e. perform the same computation) without textual similarity.
\end{description}
These types yield basic insights into the vague term of \emph{similarity} in the code clone definition.
Code fragments can be similar based on their textual representation (Type 1-3) or can have similar functionality without textual similarities (Type 4).
Type 1 and 2 clones are the focus in this paper.
Code clone types also characterize the accepted difference between code fragments participating in a clone, and they further define capability levels of detection tools.

\subsection{Clone Tools}\label{sec:existing tools}
Clone detection tools can be categorized into detection, analysis and management tools, which are often integrated into quality management platforms.
Detection tools find code clones; the results are then filtered, visualized and categorized by means of analysis tools.
Management tools track existing clones and their evolution to make them an integral part of the quality management process.

Detection tools can be basically categorized into text, token, tree, graph, metrics and model-based tools or hybrid approaches \cite{Rattan2013, Bellon2007}.
Text-based detectors use string-matching algorithms to find similar source code parts.
Token-based methods leverage lexical analysis to extract token sequences fed into a suffix-tree/array to discover clones.
Tree-based tools expose the abstract syntax tree to apply tree similarity algorithms.
Each approach has advantages and disadvantages that often limit their capabilities in detecting certain clone types.
Text-based tools can only detect Type 1 clones and by using language dependent normalizations their capabilities improve up to Type 3 clones.
Tree-based tools use computational intensive algorithms but can detect clones up to Type 3.
To summarize, detection tools are the basis of clone detection and differ in their algorithmic interpretation of source code, which ultimately affects their capabilities.
A detailed overview of detection tools is given by Bellon et al. \cite{Bellon2007}, Koschke \cite{Koschke}, and Rattan et al. \cite{Rattan2013}.

Clone analysis is concerned with filtration, visualization, and categorization of clones and is often tightly coupled with clone management.
Common visualizations are tree maps and scatter plots \cite{Ducasse, Rieger2004, Asaduzzaman2011, Ueda2002, Tairas2006}, but also parallel plots \cite{Higo2005} are used.
A tree map uses interactive tiles that reflect the directory hierarchy colored according to their duplication intensity.
Scatter plots enumerate files along the x and y-axis where each data point reflects a duplication relationship.
The combination of both provides insight into the clone relationship between but also the clone intensity within files.
Filtering and ranking of clones help to organize the typically large result sets of detectors.
This is done manually in conjunction with visualizations or automatically based on metrics and predefined criteria.
For instance, Gemini \cite{Ueda2002} or CLICS \cite{Kapser2004, Kapser2005} are tools that make use of metrics and filtering criteria to provide the most relevant subset of clones.
Categorization sorts clones into views such that the inspection can be focused on a specific task.
These views may be related to the location (\textit{Same File}, \textit{Same Directory}, etc.), the region (\textit{Function to Function Clones}, \textit{Macro Clones}, etc.), or the block classification (\textit{Initialization Clones}, \textit{Loop Clones}, etc.) \cite{Kapser2004} of the code fragments.

Management tools help to track the clones such that they can be actively incorporated into quality assessments and architectural decision processes, but also to evaluate their evolution.
This is especially important with the increased demand of modularization in the machine and plant industry \cite{Li2012}.
Clones are often deliberately introduced as light weight variability mechanism, hence they exceed typically one product life-cycle.
One way to manage these clones is, for example, CloneTracker \cite{Duala-Ekoko}.
It builds a model of the tracked clones and provides notifications if cloned code is changed or edited simultaneously.
Another example is ECCO, Extraction and Composition for Clone-and-Own \cite{Fischer2014}.
It uses fork clones in conjunction with a feature model to build a proper Software Product Line (SPL).
This allows active reuse of fork clones as they are transformed into a well-defined corpus of reusable and combinable modules.

\subsection{Related Work}\label{sec:related_work}
Code clones are well investigated by the research community resulting in a good understanding why source code is copied.
Roy and Cordy \cite{KumarRoy2007} presented a comprehensive overview of reasons for cloning extracted from various publications.
For instance, Kim et al. \cite{Kim} conducted an ethnographic study on the code clone behavior of software developers by recording the file changes.
Not only language limitations forced the developers to copy code, but developers actively used the copy and paste history to determine the abstractions within their system.
Another example, given by Kapser and Godfrey \cite{Kapser2008}, describes several different forking patterns in which large proportions of the system are copied in order to enable software ports, specific hardware implementations or (experimental) variants.

These reasons indicate -- in contrast to the incentive earlier publications give \cite{Baker, Baxter, Ducasse, Kamiya, Kontogiannis1996, Mayrand1996, Lozano2007, Juergens2009} -- that code clones are not universally bad or result of unskilled programmers.
In fact, follow-up publications found quite the contrary \cite{Monden2002, Kim, Kapser2008, Krinke2008, Krinke2011, Rahman2012, Selim2010, Bettenburg2012, Gode2011}, as Rattan \cite{Rattan2013} reported, especially with respect to the stability and faults caused by code clones.
Possible advantages of clones during development activities are risk avoidance \cite{KumarRoy2007}, architectural improvements \cite{Kapser2008}, performance improvements (e.g., loop unfolding, reduced call overhead) and improved code stability \cite{Rahman2012, Bettenburg2012, Krinke2008}.
Interestingly many found advantages also show up as disadvantage indicating that measuring the impact of clones is a non-trivial task.
Concluding, it is clear that it depends on more than just whether code is duplicated to make statements about the quality of a system.
Nevertheless, awareness and suitable methods to track and process clones are recommended, so that benefits of cloning can be leveraged and drawbacks can be mitigated.

\section{Industry Context}\label{sec:industry context}

The work described in this paper was conducted with our industry partner, a large high-tech company in the domain of machinery for metal processing.
Together we analyzed a pre-release version of a machine control software system.
It consisted of modules implemented in the IEC 61131-3 language Structured Text and modules written in the C/C++ programming language.
The total size of the software system was 191 kLOC (Lines Of Code, LOC) with 157 kLOC in ST and 34 kLOC in C/C++, at the time the study was conducted.
These numbers include only the code authored by our industry partner.

The software system was part of a large industry project and had already been evolved over several iterations with an overall development history of more than two years.
In each iteration, major functional extensions were integrated, tested and stabilized. Furthermore, every iteration also included extensions that added support for different machine types and hardware variants.
It was expected that this evolution led to code clones, as existing software routines were reused for similar hardware options by following a simple forking approach.
Hence, code fragments up to entire subsystems were copied from the existing implementation to support the requirements of the new machine types or hardware variants.

\section{Code Clone Analysis}\label{sec:code clone analysis}
We analyzed the PLC software system with Simian \cite{Harris2003a}, a proprietary text-based clone detector, and evaluated a subset of the found clones.
Simian can detect Type 1 clones in all text sources but incorporates additional normalization features for several common programming languages.
These normalization features were re-implemented for Structured Text such that all languages used in the studied system (ST and C/C++) could be analyzed on the same capability level, i.e., Type 2 clones.

\begin{table*}[!t]
\footnotesize
\centering
\ra{1.3}
\begin{threeparttable}
	\caption[Simian Results]{
	Results of the Simian clone detection on the entirety of the code base including libraries and definition files.
	}
	\label{tab:simian}
	\begin{tabular}{@{}l r r r r r r@{}}
		\toprule
		Language		& Option	& Files with Clones		& Duplicated Lines 	& Duplicated Blocks  & Total Files  & Total Sig. Lines  \\
		\midrule
		\multirow{4}{*}{C/C++} 	& Default 						& 257	& 58,741		&	1,510	&	\multirow{4}{*}{372} & \multirow{4}{*}{99,538} \\
														& Identifier 					& 334	& 99,620		&	4,005	&											 & \\
														& Literal							& 274	& 62,051		&	1,828	&											 & \\
														& Identifier/Literal 	&	340	& 117,080	&	4,776	&											 & \\
		\midrule
		\multirow{4}{*}{ST} 		& Default 						& 552	& 43,697		&	4,591	&	\multirow{4}{*}{770} & 		 \multirow{4}{*}{160,132} \\
														& Identifier 					& 633	& 105,787	&	10,930	&											 & \\
														& Literal							& 558	& 57,557		&	5,179	&											 & \\
														& Identifier/Literal 	&	650	& 133,488	&	12,291	&											 & \\
		\midrule
		\bottomrule
	\end{tabular}
	\begin{tablenotes}
		\item Clone overlap is allowed
		\item Minimum number of lines = 5
	\end{tablenotes}
\end{threeparttable}
\end{table*}

The analyzed source code has in total $99\,538$ C/C++ and $160\,132$ ST significant (non-whitespace) lines distributed over $372$ C/C++ and $770$ ST files.
This includes C/C++ and ST libraries as header files.
The source base contains multiple variants of the system for the different machine types, consequently, large portions of the code are very similar, leading to many clones.
Many of these clones are deliberately introduced and manually managed to simplify the product line aspect of the development process.
Table \ref{tab:simian} contains the number of duplicated lines, blocks, and files found by the detector.
The clone analysis did allow for clone overlaps in order to find partial copies of variant files while simultaneously allowing full copies.
The number of duplicated lines is strongly dependent on the minimum number of lines a clone is allowed to have, which was 5 lines throughout the study.
This setting is already fairly low but was chosen with the subsequent study in mind.

\section{Code Clone Study}\label{sec:code clone study}
A group of experts inspected clones found during the clone analysis (Section \ref{sec:code clone analysis}) within the system of our industry partner (Section \ref{sec:industry context}).
These inspections evaluated the nature (type) of the clones as well as their relevance for refactoring in order to help to answer the following questions:
\begin{enumerate*}
  \item What natures of clones exist within the system?,
  \item Is there a difference between the natures between C/C++ and ST?,
  \item How does the tool support influence the relevance of clones?, and
  \item How does the clone selection approach influence the relevance of clones?
\end{enumerate*}

Ultimately, these questions provide the first incentive for developers of PLC software to adapt existing clone detection tools for IEC 61131-3 languages such as Structured Text.

\subsection{Study Design}
A subset of the clone detector results was selected and presented to a group of experts.
The experts inspected the clones and provided an evaluation of each clone with respect to their nature and relevance.
The responses were then analyzed and used to explore the relationships and occurrences of clones, tool support, selection approach, and languages.

\subsubsection{Controlled Variables}
The usage of clone detection tools raises some typical questions related to the tool configuration (normalizations, minimal line length, etc.), as well to the selection approach used when analyzing the discovered clones.
Each answer potentially changes the type of clones and their perceived relevance.
Therefore it is important to understand the impact of these variables when managing clones in development and maintenance projects.
This study controls for the variables programming language, tool option, and clone selection approach.

\begin{itemize}

	\item \textbf{Language:}
	This variable refers to the used programming languages and captures the heterogeneity of the system's code base, which reflects a typical setup in which multiple technologies are used in concert to solve a complex problem.
The used languages are \textit{C/C++} and \textit{Structured Text} (ST). C/C++ are widespread general purpose programming languages for ``system-near'' applications. ST is a high-level block structured language designed for PLCs defined by the IEC 61131-3 standard. Both languages are procedural and imperative. They exhibit basic similarities but nonetheless they differ in syntax and expressiveness.

	\item \textbf{Option:}
	Simian offers basic analysis capabilities that can detect Type 1 clones in any text source.
In addition it provides normalization features for several popular programming languages to support the detection of Type 2 clones.
In this study the following combinations of options were used: \textit{Text} (source code is interpreted as normal text), \textit{Identifier} (identifiers are normalized to a common symbol), \textit{Literal} (literals are normalized to a common symbol), and \textit{Identifier + Literal} (identifiers and literals are normalized to a common symbol).

	\item \textbf{Selection:}
	Ranking and filtering of clones is used to cope with the usually large result sets.
The (confounding) selection variable reflects this behavior with the following common clone selection approaches: \textit{Random} (clones are selected randomly),
\textit{Lines} (clones are selected in ascending order according to the number of lines they span), \textit{Blocks} (clones are selected in ascending order according to the number of blocks they include).

\end{itemize}

\begin{table*}[!t]
\footnotesize
\centering
\ra{1.3}
\begin{threeparttable}
	\caption[ICC]{
		Intraclass Correlation Coefficient (ICC) of the expert responses measured by a two-way model with a fixed set of $k$ raters.
	}
	\label{tab:irr}
	\begin{tabular}{@{}l r r r r r r r r@{}}
		\toprule
				 				& 	 		& 	 		& \multicolumn{2}{c}{95\% Confidence Interval} & \multicolumn{4}{c}{F Test} \\
		\cmidrule{4-9}
		Response		& Type	& ICC		& Lower Bound 	& Upper Bound & Value 	& df1 & df2 & Sig \\
		\midrule
		Aspect 			& ICC3k & 0.819 & 0.789 				& 0.845 			& 5.528		& 479 & 958 & 0 	\\
		Logical			& ICC3k & 0.902 & 0.886 				& 0.916 			& 10.193	& 479 & 958 & 0 	\\
		Structural	& ICC3k & 0.962 & 0.955 				& 0.967 			& 26.080	& 479 & 958 & 0 	\\
		Syntactical	& ICC3k & 0.505 & 0.423 				& 0.577 			& 2.019		& 479 & 958 & 0 	\\
		Relevance		& ICC3k & 0.800 & 0.767 				& 0.829 			& 4.999		& 479 & 958 & 0 	\\
		\midrule
		\bottomrule
	\end{tabular}
	\begin{tablenotes}
		\item Number of subjects = 480
		\item Number of raters = 3\\
		\item Two-way consistency averaged-measures ICC
	\end{tablenotes}
\end{threeparttable}
\end{table*}

\subsubsection{Response Variables}
Each clone was evaluated,
\begin{enumerate*}
  \item whether it is relevant for refactoring and,
  \item to which degree it associates to the four natures: Aspect, Structural, Syntactical or Logical.
\end{enumerate*}
The resulting five response variables are given by a 5-point symmetric Likert scale ranging from \textit{Strongly disagree} to \textit{Strongly agree} with the neutral mid-point \textit{Neither agree nor disagree}.
This evaluation scheme is based on the findings of Walenstein et al. \cite{Walenstein2003} that human raters do not agree on whether a clone should be refactored or not as different developers have different emphasizes.
The Likert scale mitigates this issue by avoiding a binary decision and providing different levels of association and disassociation.
Each response is evenly mapped onto a scale between $-1$ and $1$ and averaged through all raters.
This results in interval scale data with respect to the raters but also to the number of clones inspected within each group enabling the usage of standard statistical methods \cite{Norman2010}.

\begin{itemize}

\item \textbf{Aspect:}
Clones of this nature contain statements related to cross-cutting concerns, e.g., debugging, logging, permission and authentication, data monitoring, etc.
These clones are often unavoidable and cannot be removed with common clone refactoring strategies.
Aspect Oriented Programming (AOP) frameworks are a solution to these clones and a general review on AOP methods is given by Kurdi \cite{Kurdi2013}, while Bengtsson \cite{Bengtsson2013} describes an approach specialized for IEC61131-3.

\item \textbf{Logical:}
Code fragments of logical nature describe an algorithmic unit fulfilling a specific task.
They contain a dense occurrence of computations and operations on data structures nested within control flow constructs.

\item \textbf{Structural:}
Code fragments are of structural nature if they exhibit many definitions and initializations.
They build up the structure of a software system. Typical examples are class, struct or variable definitions or initializations in header files or global constant definition files.

\item \textbf{Syntactical:}
Clones of syntactical nature are the result of text-based detectors that do not interpret whitespace or syntactical symbols (braces, brackets, etc.).
For example, series of closing braces belonging to deeply nested control flow constructs may be detected as a clone by a text-based detector.

\item \textbf{Relevance:}
Relevance captures the likelihood that an expert would issue a refactoring of a particular clone in a general maintenance scenario.
It reflects the typical true and false positive classification but avoids the forced binary decision.

\end{itemize}

\subsection{Procedures}

A subset of the found clones from Section \ref{sec:code clone analysis} was selected and presented to three experts.
Each expert was briefed in the meaning of the response variables.
Each rater was free to evaluate the clones on his own pace and the inspection sessions were done self-managed.
Each of the experts had a very strong background in software engineering.
The average experience of the experts was $11.33$ years ($\text{SD} = 4.04$ years).

Evaluation procedures computed the Inter-Rater Reliability (IRR) to quantify consistency and agreement among experts.
Further, a linear model was fitted to expose relationships between the relevance of clones and the other variables.
Finally, a set of hypothesis tests were conducted to give a further incentive on whether tool support specific for IEC 61131-3 languages are justified.

\subsection{Evaluation}

Overall 480 clones distributed over 32 groups (2 languages $\times$ 4 options $\times$ 4 selection approaches) with each containing 15 clones were inspected.
This results in a total of 1440 inspections (480 clones $\times$ 3 raters) conducted by the experts. A two-way, consistency, averaged, Intraclass Correlation Coefficient (ICC) measure \cite{McGraw1996} was used to assess the reliability of the 1440 inspections with respect to the nature and relevance.
Results within Table \ref{tab:irr} show that there was a high degree of agreement among the expert ratings over the 480 clones.
The consistency was excellent (Cicchetti interpretation guidelines \cite{Cicchetti1994}), except for the Syntactical nature only being fair ($ICC3k_{Syntactical} = 0.505$).
Given the high ICC, a minimal amount of measurement error was introduced by the experts affecting the power of subsequent analysis.
Ratings on the syntactical nature were deemed too erroneous therefore excluded from further analysis.

\begin{figure*}[!t]
  \centering
  \includegraphics[width = .7\textwidth]{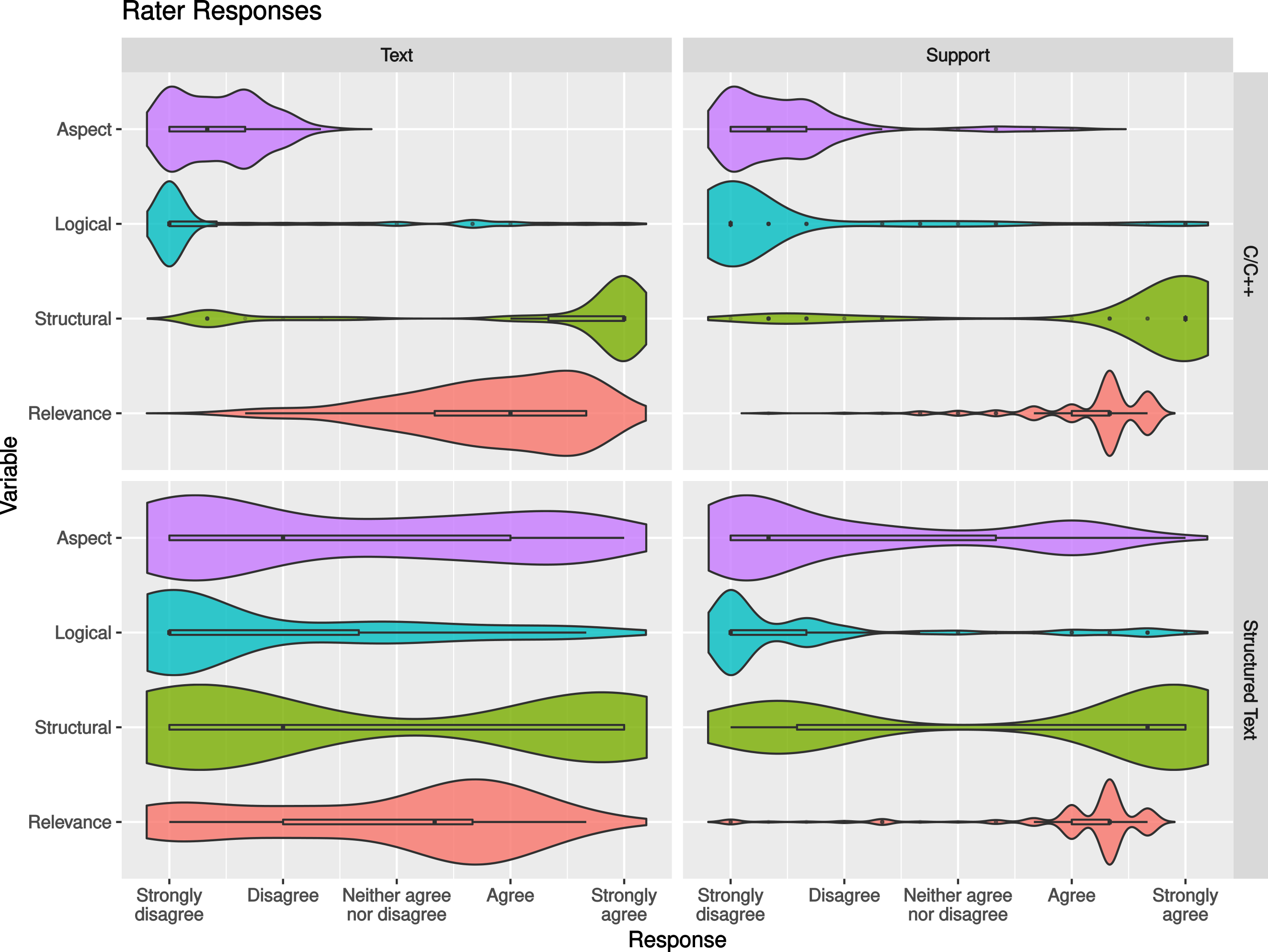}
  \caption{
  Responses with respect to the languages and natures.
  For easier interpretation the plot uses the Likert labels on the response axis (x-axis) although being values between -1 and 1.
  The Support facet includes responses from \textit{Identifier, Literal} and \textit{Identifier/Literal} option.
  }
  \label{fig:responses}
\end{figure*}

Figure \ref{fig:responses} shows the expert averaged inspection result distributions for the natures \textit{Aspect}, \textit{Logical} and \textit{Structural} as well as for \textit{Relevance}.
The left facet captures inspections of clones detected only via the \textit{Text} mode of the detector.
On the right facet are inspections of clones detected with additional normalizations (\textit{Support}) in place, i.e., normalization of identifiers, literals or both.
The distributions indicate that the experts had a clear idea whether a clone is positively associated with a nature or not.
This can be seen by the slim bellies in the neutral region of the response scale.

Clones are not associated with the \textit{Aspect} nature for C/C++ in the text mode.
This slightly changes with the usage of normalization support indicated by the longer tail of the violin shape.
ST clones are stronger associated with the \textit{Aspect} nature indicted by the third quartile reaching beyond the neutral region of the response scale.
However, the central tendency is still in the disagree range for both detector capability modes.
\textit{Logical} clones are scarce and the central tendency for C/C++ and ST are both in the strongly disagree area.
The outliers indicate the few clones that contain algorithmic content.
Most \textit{Logical} clones were found in the text mode for ST, nevertheless, the tendency is still towards a disassociation.
Many clones found in C/C++ are of \textit{Structural} nature indicated by the median located at strongly agree.
The normalization support increases the number of \textit{Structural} clones even more.
For C/C++ the first quartile moves towards the strongly agree region, for ST a shift from dissociation to an association in the central tendency of the responses is measured.

\textit{Relevance} is slightly worse for ST compared to C/C++ as the median shows, nevertheless using the additional support removes this offset and places both into the strong positive range.

\subsubsection{Linear Model}

A multivariate linear regression was calculated to predict \textit{Relevance} based on \textit{Language}, \textit{Option}, \textit{Selection} but also through the natures \textit{Logical} and \textit{Structural}.
(Note: \textit{Syntactical} has been excluded because of an insufficient reliability of the ratings and \textit{Aspect} did not reach significance.)
The maximum positive response of \textit{Relevance} is $1$ for a perfect association, $0$ for neutral and $-1$ for a perfect disassociation.
A significant regression equation was found ($F(28, 451) = 40.04, p < 2.2 \cdot 10^{-16}$), with an $R^2$ of $.713$.
Interactions between \textit{Language} and \textit{Option}, \textit{Option} and \textit{Selection}, and between \textit{Selection} and the included natures were significant.
Regression residuals show acceptable departures from normality and parallel lines as a pattern.
Patterns were expected because of the Likert scale being averaged by only three raters.
An unacceptable variation in the variances was detected, therefore heteroscedasticity corrected hypothesis tests were conducted.

The linear model shows a strong positive logarithmic relationship to the number of lines a clone spans, increasing its relevance by $0.1$ for each magnitude in lines.
Clones from ST have a lower base relevance compared to C/C++ clones ($-0.22$) but strong positive significant interactions ($0.18 - 0.26$) with the different options.
Similar, interactions that constitute blocks and normalizations in which identifiers are normalized (Identifier, Identifier+Literal), are positive significant with estimates between $0.29-0.37$.
All these coefficients indicate combinations of options, selection methods and languages that greatly increase the relevance of clones.
In terms of nature, there were strong significant coefficients that represent the interaction between \textit{Structural} or \textit{Logical} with the (between file) blocks selection method.

\subsubsection{Statistical Tests}

\begin{table*}[!t]
\footnotesize
\centering
\ra{1.3}
\begin{threeparttable}
	\caption[Planned Comparison]{
		Significance tests of specific contrasts with respect to the depended variable Relevance.
	}
	\label{tab:contrast tests}
	\begin{tabular}{@{}l r r r r r r r r r@{}}
		\toprule
				 & \multicolumn{3}{c}{Contrast} & & & \multicolumn{2}{c}{95\% Confidence Interval} & & \\
		\cmidrule{2-4} \cmidrule{7-8}
				 & Language 	& Option 					& Selection										& Mean Diff.				& Std. Error 	& Lower Bound			& Upper Bound				& Sig. 									&					\\
		\midrule
		1    & Language 	& Text - Support	& Selection										& -0.225						& 0.034				& -0.316					& -0.135						& $4.82 \cdot 10^{-10}$	& ***		\\
		2    & C/C++			& Text - Support	& Selection										& -0.116						& 0.035				& -0.209					& -0.023						& $0.005$							  & **		\\
		3    & ST					& Text - Support	& Selection										& -0.335						& 0.058				& -0.488					& -0.182						& $6.78 \cdot 10^{-8}$  & ***   \\
		4    & Language		& Option					& Random - (Blocks, Lines)		&  0.063						& 0.027				& -0.011					&  0.136						& $0.074$								&	.			\\
		5    & Language  	& Option					& Lines - Blocks							&  0.007						& 0.051				& -0.128					&  0.143						& $0.985$								&				\\
		6    & C/C++ - ST	& Option					& Selection										&  0.052						& 0.031				& -0.030					&  0.136						& $0.202$								&				\\
		7    & C/C++ - ST	& Text						& Selection										&  0.217						& 0.067				&  0.040					&  0.394						& $0.005$ 							& **		\\
		8    & C/C++ - ST	& Support					& Selection										& -0.002						& 0.030				& -0.083					&  0.079						& $0.986$								& 			\\
		\midrule
		\bottomrule
	\end{tabular}
	\begin{tablenotes}
		\item Sig. codes: 0 '***' 0.001 '**' 0.01 '*' 0.05 '.' 0.1 ' ' 1
		\item Adjusted $p$ values -- free method
	\end{tablenotes}
\end{threeparttable}
\end{table*}

The planned tests investigated whether there is a statistically significant effect in \textit{Relevance} between different groups of clones.
Contrasts measure effects within and between languages given the text mode and the average of all normalizations but also the effects of selection methods.
Table \ref{tab:contrast tests} contains contrasts and their respective hypothesis tests with \textit{free} \cite{Westfall2011} adjusted $p$ values that account for multiple comparisons.
Test 2 and 3 compare the differences between text mode and additional support within the two languages where both reach significance, although ST with a larger effect.
No marginal significant effect between random selection, and block or line oriented selection approaches could be found (Test 4).
However, the low $p$-value and the existence of interactions indicate that significant effects between specific levels of the variables are present.
Tests 6, 7 and 8 represent between language tests with no marginal effect (Test 6).
Test 7 shows a significant difference  of clones between the two languages given that they were detected with the text mode.
Test 8 shows that this significance is not found if clones are detected with additional normalizations.

\subsection{Interpretation}

The results show that the tool support has a positive effect on the relevance of clones.
This can be seen in Figure \ref{fig:responses} where the median and first quartile moves into the strong positive region, but also in the hypothesis tests.
This positive effect is given in the within language tests (Test 2 \& 3) but also in between languages tests where the initial significant difference is removed by the additional normalization features.
The between language effect is most likely caused by the header files (.h-files) of C/C++ that introduce more structural duplicates, which in hindsight are often relevant (linear model coefficient).

The selection approach does not influence the relevance of clones on average.
However, there were positive effects associated with selection methods based on the number of blocks that are shared between files.

Most clones are of \textit{Structural} nature and the usage of normalizations increases their total proportions making them more likely to be encountered.
\textit{Logical} clones are inversely proportional to structural clones and therefore less often encountered with normalizations.
Clones of \textit{Aspect} nature are mostly found with a low minimum line count of clones but remain strongly dependent on the application context.
The nature of clones between the languages is fairly similar with mostly structural clones and some logical clones.
ST code contained more aspect clones nevertheless these are less prevalent if normalizations are used.

\section{Threats to Validity}
The study faces threats to validity that might reduce the power of the analysis.
First, the generalization of results is limited because only one software system was analyzed.
However, the system is a real-world example and the applied development approach can be considered representative for many other evolving software systems for industrial automation \cite{Vogel-Heuser2015a}.
Furthermore, the choice of the analysis tool and its implementation, as the experts and their specific background in software development may also have introduced a bias.
Finally, the confounding configuration problem discussed by Wang \cite{Wang} might be an issue.
We chose the minimal line count for a clone fairly low such that aspect clones are easier spotted.

\section{Summary and Conclusions}\label{sec:conclusion}
In this paper, we presented the results from the analysis of code clones in a real-world PLC software system, which has been evolved over several development iterations as part of a large industry project.
The software system contained code written in C/C++ and in IEC 61131-3 Structured Text.

We found that clones do exist in PLC software systems regardless of the applied programming language.
Awareness for clones is an important aspect of professional software development, independent whether they are viewed positive or negative.
Industry projects require support for detecting, tracking and managing clones as software systems evolve.
Similarly to previous studies \cite{Vogel-Heuser2015a}, we can also conclude that the existing tool support for PLC languages with respect to clone detection is insufficient.
Furthermore, we found that language adaptations for detectors, that enable the use of normalizations, improve the relevance of clones significantly.
This is especially true for maintenance scenarios focusing on structural deficiencies.
Concluding, companies that develop PLC systems can justify investments in clone detector adaptations.
These investments widen the range of clone detection, analysis and management tools and strengthen professional software development within the industrial automation industry.

Future work includes methodologies for efficient filtering of clones based on their nature through complexity metrics and the repetition of the study on other PLC software systems including systems from different industry partners.

{
\ifCLASSOPTIONcompsoc
  \section*{Acknowledgments}
\else
  \section*{Acknowledgment}
\fi

The research reported in this paper has been supported by the Austrian Ministry for Transport, Innovation and Technology, the Federal Ministry of Science, Research and Economy, and the Province of Upper Austria in the frame of the COMET Center SCCH (FFG \#844597).

}

\cleardoublepage
\setstretch{.89}

\bibliographystyle{IEEEtran}

\bibliography{citations}

\end{document}